\documentclass{iau}
\usepackage{graphicx}

\title[IAUS300.~~ Physical signatures of filaments and coronal holes] 
{Featuring dark coronal structures: \\
physical signatures of filaments and coronal holes for automated recognition}

\author[J.~ Palacios et al.]   
{Judith Palacios$^1$, Consuelo Cid$^1$, Elena Saiz$^1$, Yolanda Cerrato$^1$, and Antonio Guerrero$^1$}

\affiliation{$^1$Space Reseach Group--Space Weather, Physics Dpt., University of Alcal\'a \\ 
University Campus, Sciences Building, P.O. 28871, Alcal\'a de Henares, Spain \\
 email: {\tt judith.palacios@uah.es} }
\pubyear{2013}
\volume{300}  
\pagerange{xxx}
\jname{IAUS300, Nature of prominences and their role in Space Weather}
\editors{B.~Schmieder, J.~M. Malherbe \& S. Wu, eds.}
\begin{document}

\maketitle

\begin{abstract}
Filaments may be mistaken for coronal holes when observed in extreme ultraviolet (EUV) images; however, a closer and more careful look reveals that their photometric properties are different. The combination of EUV images with photospheric magnetograms shows some characteristic differences between filaments and coronal holes. We have performed analyses with 7 different SDO/AIA wavelengths (94, 131, 171, 211, 193, 304, 335~\AA) and SDO/HMI magnetograms obtained in  September 2011 and March 2012  to study coronal holes and filaments from the photometric, magnetic, and also geometric point of view, since projection effects play an important role on the aforementioned traits.

\end{abstract}

\firstsection 
\section{Introduction: coronal holes and filaments}
Dark features observed in EUV can be coronal holes (CHs) or filaments (Fs). CHs present a unipolar magnetic field. This unipolar magnetic field helps particles to escape, originating fast solar wind. 
Filaments are dark structures, with a very different magnetic field topology. Best observed in H$\alpha$, they are located mainly over neutral lines.
The goal of this study is to check the photometric properties of CHs and filaments aiming at an automated recognition purpose.

\section{Data}
For this work we have used SDO data (\cite[Pesnell  \etal\,2012]{Pesnell2012}); more precisely, EUV data from AIA instrument of the following wavelengths: 193, 211, 304, 94, 171, 94, 335~\AA, and also longitudinal magnetic field data from SDO/HMI. These datasets are incomplete due to the eclipse season at the equinoxes. The cadence for the study is 30 min. The filaments studied are quiescent, away from active regions, and the low-latitude CHs are long-lived ($>$ 10 days) in the same field of view. Datasets from 2012, March 08-19 and 2011 Sept 06-15 are studied, but only the first one will be shown. 

\section{Photometric and magnetic properties of coronal holes and filaments}
Some doubts may arise when automatically detecting dark features in AIA calibrated images. The coronal hole/filament photometric characteristics are found by studying the intensity histograms. First of all, a threshold is set for each wavelength. Then, these photometric histograms are studied in every wavelength for filaments and coronal holes. Running windows are set to follow the regions of interest across the solar disk. When necessary, the lambertian equal-area projection was used (\cite[Krista~\etal\, 2011]{Krista2011}).

\begin{figure}[h!]
\begin{center}
\includegraphics[width=2.0in ]{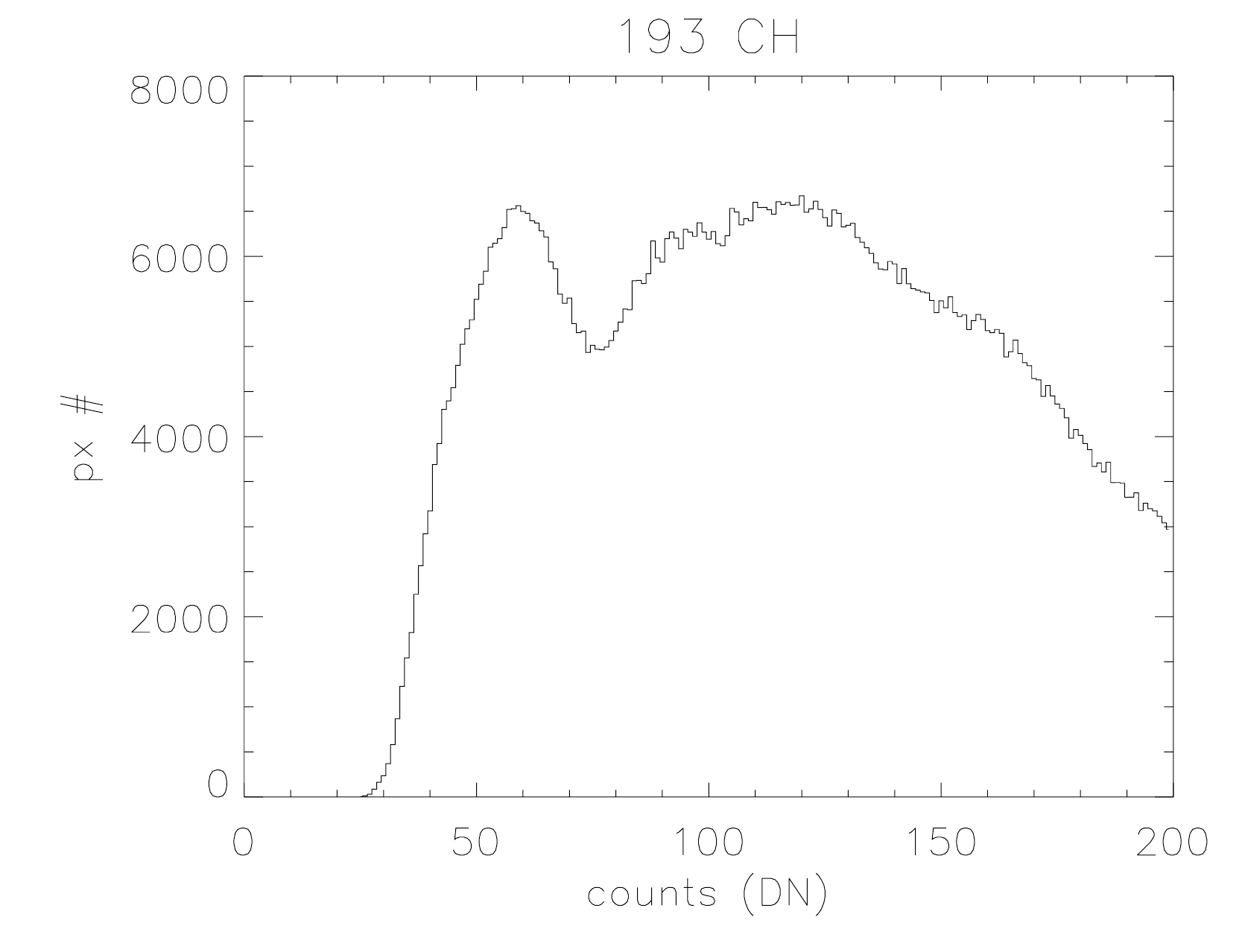} ~~
 \includegraphics[width=2.0in ]{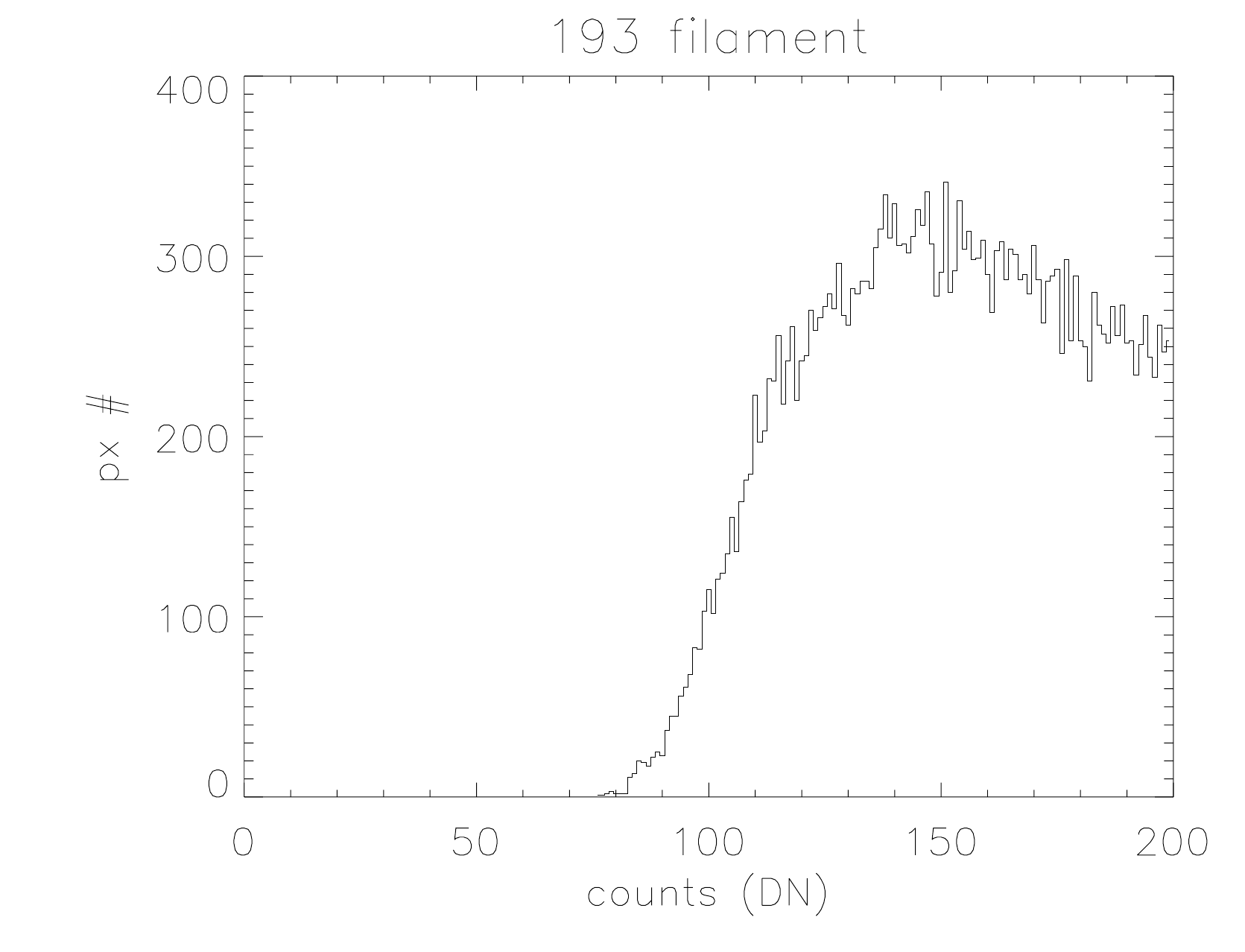} 
\caption{$\it{Left:}$ Intensity histogram of a large CH in AIA 193~\AA. X-axis shows the intensity, while the corresponding number of pixels is counted. $\it{Right:}$ Intensity histogram of a filament in AIA 193~\AA.}
\label{fig1}
\end{center}
\end{figure}

In full-disk images, the intensity distribution can mark different features: bimodal intensity distributions show the existence of AR and CHs for 195~\AA~images (\cite[Krista \& Gallagher, 2009]{KristaGallagher2009}).
In this work we show different intensity distributions inside CH and filament areas. CH intensity follows a bimodal distribution, clearly marked in 193 and 211~\AA~(Fig.\,~\ref{fig1} $\it{left}$) while filaments usually display a unimodal distribution (Fig.\,~\ref{fig1} $\it{right}$). AIA wavelengths 94~\AA~and 131~\AA~are too noisy for dark regions, since their temperature response is more adequate for flaring regions (\cite[O'Dwyer~\etal\, 2010]{O'Dwyer2010}).

Regarding the magnetic field of these features, when filaments (and CHs) are located and contoured, their areas are superimposed to HMI magnetograms to relate the EUV image to the photospheric magnetic field. The filament mean magnetic field is around -0.5~G for all wavelengths.

Besides the HMI analyses, aiming at locating neutral lines, HMI images also can be used to pinpoint areas where quiescent filaments may be located, using a segmentation method.

\section{Final remarks and conclusions}
We have studied filaments and CHs. Generally, the mean magnetic field is higher for CHs than filaments, while intensity thresholds are usually higher for filaments than for coronal holes across the solar disk. The intensity histogram profiles are also different in filaments and CHs.

\section{Acknowledgements}
The authors want to acknowledge SDO/AIA and SDO/HMI Data Science Centers and Teams; and we would like to thank funding from the Spanish project PPII10-0183-7802 from ``Junta de Comunidades de Castilla -- La Mancha'', and also from ESA, through the travel grant to participate in the IAUS300 Symposium.

\end{document}